\documentclass[showpacs,preprintnumbers,amssymb,twocolumn]{revtex4-2}
\usepackage{graphicx}
\usepackage{dcolumn}
\usepackage{color}
\usepackage{bm}
\usepackage{amsmath}
\usepackage{amssymb}
\usepackage{epsfig}
\usepackage{amsfonts}
\usepackage{lineno,hyperref}
\usepackage{array}
\usepackage{microtype}
\usepackage{float}
\usepackage{multirow}
\usepackage{adjustbox}
\usepackage[english]{babel}
\usepackage{blindtext}
\usepackage[a4paper, total={7.5in, 10in}]{geometry}

\def \a{\alpha}
\def \b{\beta}
\def \l{\lambda}

\def \g{\gamma}
\def \Ga{\Gamma}
\def \d{\delta}

\def \m{\mu}
\def \n{\nu}
\def \o{\omega}

\def \p{\partial}
\def \t{\theta}
\def \r{\rho}
\def \T{\Theta}
\def \nb{\nabla}
\def \non{\nonumber}
\def \ben{\begin{eqnarray}}
\def \een{\end{eqnarray}}
\def \G{\bar{G}}
\def \R{\bar{R}}
\def \dd{\ddot{\phi}}

\def \S{\bar{s}}
\def \D{\dot{\phi}}

\begin{document}
\title{Raychaudhuri equation in {\it{k}}-essence geometry: conditional singular and non-singular cosmological models}

\author{Surajit Das}
\altaffiliation{surajit.cbpbu20@gmail.com}
\affiliation{Department of Physics, Cooch Behar Panchanan Barma University,
Panchanan Nagar, Vivekananda Street, Cooch Behar, West Bengal, India 736101}

\author{Arijit Panda}
\altaffiliation{arijitpanda260195@gmail.com}
\affiliation{Department of Physics, Raiganj University, Raiganj, Uttar Dinajpur 733 134, West Bengal, India. $\&$\\
Department of Physics, Prabhat Kumar College, Contai, Purba Medinipur 721404, India}

\author{Goutam Manna$^{a}$ }
\altaffiliation{goutammanna.pkc@gmail.com\\
$^{a}$Corresponding author}
\affiliation{Department of Physics, Prabhat Kumar College, Contai, Purba Medinipur 721404, India}

\author{Saibal Ray}
\altaffiliation{saibal.ray@gla.ac.in}
\affiliation{Centre for Cosmology, Astrophysics and Space Science (CCASS), GLA University, Mathura 281406, Uttar Pradesh, India}

\date{Received: date / Accepted: date}

\begin{abstract}
We investigate how the Raychaudhuri equation behaves in the {\it k-}essence geometry. As far as we are concerned, both the early and current epochs of the universe are relevant to the {\it k-}essence theory.  Here, we have studied the {\it k-}essence geometry using the Dirac-Born-Infeld (DBI) variety of non-standard action. The corresponding {\it k-}essence emergent spacetime is not conformally equivalent to the usual gravitational metric. We assume that the background gravitational metric is of the Friedmann-Lemaitre-Robertson-Walker (FLRW) type in this case. We have found that both the conditional singular and non-singular cosmological models of the universe through the modified Raychaudhuri equation are possible where we have used the spacetime as the flat {\it k-}essence emergent FLRW-type. We have also addressed to the Focusing theorem and conditional caustic universe construction. These conditional effects are caused by the additional interactions that arise as a result of the coupling that exists between the gravity and the {\it k}-essence scalar field.
\end{abstract}

        
\maketitle

\section{Introduction}\label{s1}
The Raychaudhuri equation is largely viewed as one of the most elegant equations, and it has made significant advancements in the study of Einstein's theory of general relativity \cite{blau,poisson}. In 1955, Raychaudhuri derived his famous equation \cite{ray55}, which is purely geometric  having no relationship with the Einstein field equation \cite{einstein}. Later he derived the said equation with a modern approach \cite{ray57}. The Raychaudhuri equation deals with the notion of the congruence of geodesics. If $M$ be a manifold and $O\subset M$ be open then a congruence in $O$ is a family of curves such that, through each point $p\in O$, there passes precisely one curve in this family. Obviously, no two trajectories within the family can intersect each other on time evolution, failling this, the definition of congruence breaks down. Geodesic congruence is the process of tracking the pathways taken by a group of particles that do not interact as they travel across a generic curved spacetime. The goal of the Raychaudhuri equation is to investigate and study of the temporal behaviour of a congruence of geodesics as observed by a neighbouring member of the congruence of geodesics itself. So, the equation tells us, how the rate of change of the scalar expansion of a test particle relates to the gravity effect, i.e., the energy-momentum of that test particle.

The goal of the present study is to develop the Raychaudhuri equation and  study its effects in a more general form for the special case of {\it k}-essence emergent gravity \cite{vikman1}. The {\it{k}}-essence model \cite{vikman1,vikman2,picon1,picon2,picon3,scherrer,chimento1,chimento2,chimento3} is a scalar field model and the field Lagrangian is non-canonical in nature. In {\it{k}}-essence model, the kinetic energy of the field dominates over the potential energy. The non-trivial dynamical solutions of the {\it k}-essence equation of motion that set it apart from relativistic field theories with canonical kinetic components, since they both violate Lorentz invariance spontaneously and modify the metric for perturbations near these solutions.
The so-called analogue or emergent curved spacetime metric is affected by the relevant perturbations because the theoretical form of the {\it k-}essence field Lagrangian is non-standard. A form of the {\it k-}essence Lagrangian is
$L =-V (\phi)F(X)$, where $\phi$ is the {\it k}-essence scalar field and $X =\frac{1}{2} g_{\mu\nu} \nabla^{\mu} \phi \nabla^{\nu} \phi$. In the literatures \cite{gm1,gm2,gm3,gm4,gm5,Mukohyama}, the authors have used the Dirac-Born-Infeld (DBI) \cite{Born,heisenberg,Dirac,Varshney2021,Debnath2022,Born2} type non-canonical Lagrangian to study the {\it k}-essence geometry and its consequences. But Tian et al. \cite{Tian} have studied the {\it k}-essence geometry and it's effect using different types of non-canonical Lagrangian which is an arbitrary function of $\phi$ and $X$. 

Our universe is expanding faster than ever before, as we already know from the study of observations of large-scale structure, measurements of type Ia supernovae and CMBR \cite{bahcall,Riess,Perlmutter,Komatsu,planck1}. Additionally, it is well known that, a unique element known as dark energy presently dominates our universe. However, the issue of why the unusual dark energy component has such a low energy density $(\mathcal{O}(meV^4))$ in contrast to the straightforward expectation based on quantum field theory arises. The issue with the different dark energy theories is that they call for very precise tuning of the beginning energy density, which is at least a factor of 100 or more, less than the initial energy density of the matter. The scalar field model, or {\it k-}essence theory, eliminates the fine-tuning difficulty \cite{picon1} and may also provide an explanation for  the question, why acceleration occurred at such a late stage of time development. The model's ability to generate negative pressure due to its nonlinear kinetic energy linked with this scalar field model is its most crucial feature. During the radiation-dominated epoch, {\it k-}essence monitors the equation of state of the background universe. Therefore, {\it k}-essence may automatically change into an effective cosmological constant at the beginning of matter-domination to explain the current cosmic acceleration \cite{picon1}.

On the other side, if there is an early component of dark energy that becomes active around the time of matter-radiation equivalence, the Hubble tension issue may be greatly reduced \cite{sakstein}. Around the moment of matter-radiation equivalence, the early dark energy scalar couples with neutrinos and obtains a significant energy boost. Early dark energy thus emerges precisely at the moment of the equality of matter and radiation, when neutrinos become non-relativistic at the energy scale of $\mathcal{O} (eV)$. Therefore, the neutrino-assisted model, which best exemplifies how the beginning and end of early dark energy are governed by the matter-radiation transition, may be used to resolve the coincidence issue of early dark energy \cite{sakstein,poulin}. As a result, the {\it k}-essence model realises an early dark energy model with several scenarios. Therefore, the {\it k-}essence scalar field model encompasses the whole spectrum of our universe's history, from its infancy to its late-time acceleration \cite{picon1,picon2,picon3}. Despite the enigmatic cosmic context of dark energy, whose very existence may now be in some doubt \cite{Sarkar}, based on the recent data from the Planck consortium \cite{Ade}. In this setting, the {\it k}-essence concept may be used not just for the study of dark energy, but also for gravitational or geometrical aspects \cite{gm4,gm5}.

The paper is planned as follows : In Sec.~\ref{s2}, we have briefly discussed the {\it{k}}-essence emergent gravity model. In Sec.~\ref{s3}, we have derived the Raychaudhuri equation in {\it{k}}-essence emergent spacetime with the help of the theory of Ricci identity \cite{chandrasekhar}. For a scalar field $\phi(t)$, we have formulated the Raychaudhuri equation for a choice of cosmological settings i.e., in a comoving Lorentz frame. For this, we have used the background gravitational metric to be flat FLRW-type. In Sec.~\ref{s4}, we have dealt with cosmological models for a particular choice of a unit vector field along the temporal direction only. Here, under some conditions, we get three types of cosmological models, i.e., the collapsing universe, the expanding universe and the steady-state universe. Raychaudhuri already told about the collapsing universe without assuming the symmetries of the spacetime. Our model has satisfied with that, for a particular condition. On the other hand, we also get a non-singular universe model for a particular condition. As {\it{k}}-essence answered about the late time acceleration of our universe, so it is quite natural for getting a bouncing scenario of our universe instead of collapse. Except for these two models, we also get the conditional steady-state model of our universe \cite{hoyle,bg,hn1,hn2}.  On the other hand, the congruence of geodesics develops a point at which some of the geodesics from the congruence come together and this point is a singularity of the congruence, called a caustic point. But here we have also mentioned the non-singular theory as an expanding universe for initially diverging congruence of geodesics. So, finally, we have discussed the Focusing theorem and caustic formation in modified cosmology with two different casses in Sec.~\ref{s5} and end up with a brief conclusion and discussions in Sec.~\ref{s6}.

\section{{\it k}-essence emergent geometry}\label{s2}
In this section, we have presented a brief review of the  {\it k}-essence geometry. The {\it k}-essence action, where the {\it k}-essence scalar field $\phi$ has minimally coupled with the gravitational metric $g_{\mu\nu}$ \cite{vikman1,vikman2,picon1,picon2,picon3,scherrer}:
\ben
S_{k}[\phi,g_{\mu\nu}]= \int d^{4}x {\sqrt -g} L(X,\phi),
\label{1}
\een
where $L(X,\phi)$ is the {\it non-canonical} Lagrangian with $X=\frac{1}{2}g^{\mu\nu}\nabla_{\mu}\phi\nabla_{\nu}\phi$ and the form of the energy-momentum tensor is:
\ben
&T_{\mu\nu}\equiv \frac{-2}{\sqrt {-g}}\frac{\delta S_{k}}{\delta g^{\mu\nu}}=-2\frac{\partial L}{\partial g^{\mu\nu}}+g_{\mu\nu}L\nonumber\\
&=-L_{X}\nabla_{\mu}\phi\nabla_{\nu}\phi
+g_{\mu\nu}L,\label{2}
\een
with $L_{\mathrm X}= \frac{dL}{dX},~~ L_{\mathrm XX}= \frac{d^{2}L}{dX^{2}},
~~L_{\mathrm\phi}=\frac{dL}{d\phi}$ and  
$\nabla_{\mu}$ is the covariant derivative defined with respect to the gravitational metric $g_{\mu\nu}$.

The corresponding scalar field equation of motion (EOM) is
\ben
-\frac{1}{\sqrt {-g}}\frac{\delta S_{k}}{\delta \phi}= \tilde{G}^{\mu\nu}\nabla_{\mu}\nabla_{\nu}\phi +2XL_{X\phi}-L_{\phi}=0,
\label{3}
\een
where  
\ben
\tilde{G}^{\mu\nu}\equiv \frac{c_{s}}{L_{X}^{2}}[L_{X} g^{\mu\nu} + L_{XX} \nabla ^{\mu}\phi\nabla^{\nu}\phi],
\label{4}
\een
with $1+ \frac{2X  L_{XX}}{L_{X}} > 0$ and $c_s^{2}(X,\phi)\equiv{(1+2X\frac{L_{XX}}
{L_{X}})^{-1}}$.

The inverse metric is
\ben G_{\mu\nu}=\frac{L_{X}}{c_{s}}[g_{\mu\nu}-{c_{s}^{2}}\frac{L_{XX}}{L_{X}}\nabla_{\mu}\phi\nabla_{\nu}\phi].
\label{5}
\een

After taking a conformal transformation \cite{gm1,gm2} $\bar G_{\mu\nu}\equiv \frac{c_{s}}{L_{X}}G_{\mu\nu}$ we have
\ben 
\bar G_{\mu\nu}
={g_{\mu\nu}-{{L_{XX}}\over {L_{X}+2XL_{XX}}}\nabla_{\mu}\phi\nabla_{\nu}\phi}.
\label{6}
\een
	
It is to be noted that $L_{X}\neq 0$ for a positive definite $c_{s}^{2}$, has to be hold to make Eqs. (4)--(6) physically meaningful. From Eq. (\ref{6}) we can say that, the emergent gravity metric  $\bar G_{\mu\nu}$ is not conformally equivalent to the usual gravitational metric $g_{\mu\nu}$ for this type of non-trivial spacetime. Again, if we assume the $L$ is not depended on $\phi$ explicitly then the EOM (\ref{3}) is reduces to:
\ben
-{1\over \sqrt {-g}}{\delta S_{k}\over \delta \phi}
= \bar G^{\mu\nu}\nabla_{\mu}\nabla_{\nu}\phi=0.
\label{7}
\een

Now we take a special type of non-canonical Lagrangian which is Dirac-Born-Infeld (DBI) Lagrangian \cite{gm1,gm2,gm3,gm4,gm5,Mukohyama,Born,heisenberg,Dirac,Born2}:
\ben
L(X,\phi)\equiv L(X)= 1-V\sqrt{1-2X}, \label{8}
\een
where $V(\phi)\equiv V$ is a constant potential and which is very small compared to the kinetic energy of the {\it k-}essence scalar field.

Now we discuss the explanation of the choice of the DBI type non-canonical Lagrangian (\ref{8}). DBI theory \cite{Born} was first proposed to eradicate the problems in solving the infinite-self energy by quantum theory. But the most promising gift given by the DBI-type model to cosmology is that it can explain the inflationary phenomenon based on string theory \cite{Guth,Linde,Pedram,Evans,Gwyn}. Basically, this theory provides the field for inflation using the motion of the D-brane in the 6-dimensional compact submanifold of spacetime \cite{Zhang}. D-branes, abbreviation for Dirichlet membrane, are a class of extended objects upon which open
strings can end with Dirichlet boundary conditions. Moreover, the DBI-type action associated with a non-canonical kinetic term can produce inflation with steep potential instead of slow-roll inflation \cite{Zhang}. Chimento et. al. \cite{Chimento} has gone further and showed that one can show late time acceleration with the purely kinetic DBI action. Furthermore, Manna et al. \cite{gm4,gm5}  have utilised this form of DBI Lagrangian in their literatures from a strictly gravitational viewpoint, not in the context of dark energy.
Therefore, we were motivated to use the DBI-type Lagrangian in our theory. 
   
From the definition of Euler-Lagrange equation, we know that the form of the usual Lagrangian is not unique \cite{Rana}. Additionally, in the special relativistic dynamics, the Lagrangian is not of the form of ($T-V$), rather the free particle Lagrangian in special relativistic dynamics can be compared with the form DBI-type Lagrangian $L(X)=1-V\sqrt{1-2X}$, where $X=\frac{1}{2} g^{\mu\nu}\partial_{\mu}\phi\partial_{\nu}\phi$  corresponds to the kinetic energy term. According to Goldstein and Rana \cite{Goldstein,Rana}, we can say that the non-canonical form of Lagrangian is the general one, which leads to the canonical form for a specific condition. We can be extract the canonical Lagrangian from the DBI type non-canonical Lagrangian (\ref{8}) with an example in the following way: 
In an FLRW background, the DBI Lagrangian for a homogeneous {\it k-}essence field, i.e., $\phi(r,t)=\phi(t)$, with constant potential $V$ and $\dot\phi^{2} < 1$ takes the form
$L(X,\phi)= 1-V\sqrt{1-2X}\simeq 1-V+V[\frac{1}{2}g^{00}\p_{0}\phi \p_{0}\phi]
=\frac{V}{2}[\dot\phi^{2}]-(V- 1)$,
with $\dot\phi^{2}>>V$.
In this case, we have the Lagrangian written in a form reminiscent of canonical Lagrangian 
i.e. $L\equiv T-V\equiv kinetic~ energy - potential~ energy$,  where $T$ is quadratic in the time derivatives and everything else is potential. 

On the other hand, Babichev \cite{Babi} has shown that for any non-linear pure {\it k}-essence model, we can choose $L\equiv L(X)$. Particularly, the form of the Lagrangian $L(X)=K_1-K_1 \sqrt{1-2X}$ has been deduced in \cite{Mukohyama}, where $K_1$ is a constant. There is also a physical reason behind the choice of $L(X)$. As the {\it k-}essence demands for the domination of kinetic energy over
potential energy, we can neglect the effect of potential $(V)$ and consider it as a constant quantity and choose the Lagrangian to be the explicit function of $X$ only, not the explicit function of $\phi$. The consideration of scalar field model is also intriguing in recent days because it can explain dark matter, dark energy and inflation. It is to be noted that, the minimally coupled scalar field is the first and easiest model of inflation in the slow roll approximation \cite{Quaglia}. Davari et. al. \cite{Davari} in their paper studied the cosmological constrains on minimally and non-minimally coupled scalar field models and made a comparison considering the latest cosmological data. They concluded that the minimally coupled scalar field is preferable over the non-minimally coupled scalar field by a small margin .

It should be mentioned that in the {\it k}-essence geometry, the kinetic energy of the {\it k}-essence scalar field dominates over the potential energy of that field and for this Lagrangian (\ref{8}) the value of the $c_{s}^{2}$ is $(1-2X)$. Therefore, the effective emergent metric (\ref{6}) is:
\ben
\bar G_{\mu\nu}= g_{\mu\nu} - \nabla_{\mu}\phi\nabla_{\nu}\phi= g_{\mu\nu} - \partial_{\mu}\phi\partial_{\nu}\phi,
\label{9}
\een
since $\phi$ is a scalar. 

Following \cite{gm1,gm2} the connection coefficients of the {\it k}-essence emergent geometry is given by
\ben
\bar\Gamma ^{\alpha}_{\mu\nu}=\Gamma ^{\alpha}_{\mu\nu} -\frac {1}{2(1-2X)}\Big[\delta^{\alpha}_{\mu}\partial_{\nu}
+ \delta^{\alpha}_{\nu}\partial_{\mu}\Big]X,
\label{10}
\een
which is different from the usual connection coefficients ($\Gamma ^{\alpha}_{\mu\nu}$). 

Therefore the corresponding geodesic equation of this geometry also changed as:
\ben
\frac {d^{2}x^{\alpha}}{d\l^{2}} +  \bar\Gamma ^{\alpha}_{\mu\nu}\frac {dx^{\mu}}{d\l}\frac {dx^{\nu}}{d\l}=0, \label{11}
\een
where $\l$ is an affine parameter.

\section{Derivation of Raychaudhuri equation in {\it{k}}-essence for timelike geodesic}\label{s3}

We define the covariant derivative $D_{\a}$ for a timelike vector field $v^{\b}$ associated with the emergent metric tensor $\bar{G}_{\m\n}$ as: 
\ben
D_{\a} v^{\b}=\p_{\a}v^{\b}+\bar{\Ga}^{\beta}_{\alpha\mu}v^{\m}.\label{12}
\een

From Eq. (\ref{12}), we also have
\ben
D_{\a}v^\b=\nb_\a v^\b-\frac{1}{2(1-2X)}\left(\d^\b_\a v^\m\p_\m X+v^\b\p_\a X\right).\label{13}
\een

Now we can define the commutation relationship of the covariant derivatives \cite{blau,poisson,chandrasekhar} in the emergent geometry as:
\ben
[D_\a\, ,D_\b]v^\m=\bar{R}^\m_{\g\a\b}v^\g, \label{14}
\een
where $\bar{R}^\m_{\g\a\b}$ is the Riemann tensor associated with emergent spacetime.

Contracting Eq.(\ref{14}) with $v^\b$ and also over the indices $\m,\a$, we obtain
\ben
D_\a \dot{v}^\a-(D_\a v^\b)(D_\b v^\a)-v^\b D_\b(D_\a v^\a)=\bar{R}_{\g\b}v^\g v^\b, \label{15}
\een
where $\bar{R}_{\g\b}$ is the Ricci tensor and $\dot{v}^\a\equiv v^\b D_\b v^\a$ is the acceleration term associated with the emergent geometry. We call $D_\a v^\a\equiv \T$ is the scalar expansion of this geometry. For affinely parametrised geodesic equation in this background as $\dot{v}^\a=0$, we get the Raychaudhuri equation in the {\it{k}}-essence emergent spacetime as:
\ben
\frac{d\T}{d\S}+(D_\a v^\b)(D_\b v^\a)=-\bar{R}_{\g\b}v^\g v^\b.
\label{16}
\een

Now we derive the some quantities:
\ben
&&(D_\a v^\b)(D_\b v^\a)=(\nb_\a v^\b)(\nb_\b v^\a)-\frac{\t}{(1-2X)}v^\m\p_\m X\nonumber\\&&+\frac{7(v^\m\p_\m X)^{2}}{4{(1-2X)}^2}, 
\label{17}
\een
and
\ben
(\nb_\b v^\a)(\nb_\a v^\b)
= 2{\sigma}^2-2{\o}^2+\frac{1}{3}\t^2,
\label{18}
\een
where we consider affinely parametrised geodesic equation for the usual gravitational case and we set \cite{blau,poisson,akr} $v^\a\nb_\a v^\b=0$, $\t=\nb_\a v^\a$ is the scalar expansion, symmetric shear $\sigma_{\a\b}=\frac{1}{2}\Big(\nb_\b v_\a+\nb_\a v_\b\Big)-\frac{1}{3}\t h_{\a\b}$,  
antisymmetric rotation $\o_{\a\b}=\frac{1}{2}\Big(\nb_\b v_\a-\nb_\a v_\b\Big)$, $2\sigma^{2}=\sigma_{\a\b}\sigma^{\a\b}$, $2\o^{2}=\o_{\a\b}\o^{\a\b}$ and 
 $\nb_\b v_\a=\sigma_{\a\b}+\o_{\a\b}+\frac{1}{3}\t h_{\a\b}$ with the three dimensional hypersurface metric $h_{\a\b}=g_{\a\b}-v_\a v_\b$.

Using Eqs. (\ref{17}) and (\ref{18}) in Eq. (\ref{16}) we have
\ben
&&\frac{d\T}{d\S}+\Big(2{\sigma}^2-2{\o}^2+\frac{1}{3}\t^2\Big)-\frac{\t}{(1-2X)}v^\m\p_\m X\nonumber\\&&+\frac{7(v^\m\p_\m X)^{2}}{4{(1-2X)}^2} =-\R_{\g\b}v^\g v^\b.\label{19}
\een

Now, we consider the {\it k}-essence scalar field to be function of time only ($\phi(x^{i},t)\equiv \phi(t)$) \cite{gm4,gm5} since in the {\it k}-essence geometry the dynamical solutions of the {\it k}-essence scalar fields spontaneously break the Lorentz symmetry \cite{vikman1,vikman2}. So, $X=\frac{1}{2}g^{\mu\nu}\nabla_{\mu}\phi\nabla_{\nu}\phi=\frac{\dot\phi^{2}}{2}$ with $\dot\phi=\frac{\p\phi}{\p t}$.

For cosmological setting, we choose the matter to move along geodesics. So one has $v^\a=(1,0,0,0)$ in a comoving Lorentz coordinate. So from Eq. \ref{19} we get
\ben
&&\frac{d\T}{d\S}+\Big(2{\sigma}^2-2{\o}^2+\frac{1}{3}\t^2\Big)-\frac{\t~\dot{\phi}~\dd}{(1-\dot{\phi}^2)}+\frac{7\dot{\phi}^2~{\dd}^2}{4{(1-\dot{\phi}^2)}^2}\nonumber\\&&=-\R_{00}.
\label{20}
\een

Considering the background gravitational metric to be flat Friedmann-Lemaitre-Robertson-Walker (FLRW), then from Eq. (\ref{9}) we have the corresponding emergent gravity line element as \cite{Panda1,gm7,Panda2}
\ben
dS^{2}=(1-\dot\phi^{2})dt^{2}-a^{2}(t)\sum_{i=1}^{3} (dx^{i})^{2},
\label{21}
\een
where $a(t)$ is scale factor and from the EOM (\ref{7}) we have the relationship between the Hubble parameter ($H(t)$) and the {\it k}-essence scalar field \cite{Panda1,gm7,Panda2} as
\ben
\frac{\dot a}{a}=H(t)=-\frac{\ddot\phi}{\dot\phi(1-\dot\phi^{2})}.
\label{22}
\een

It should be mentioned that from Eq. (\ref{21}), the value of $\dot{\phi}^2$ in between $0$ and $1$, i.e., $0<\dot\phi^{2}< 1$, it can be clarify below. If $\dot\phi^{2}> 1$ the signature of the metric (\ref{21}) is illdefined. Also, $\dot\phi^{2}\neq 0$ condition should be hold true to apply the {\it k}-essence theory. 

From the flat emergent FLRW metric (\ref{21}) with the definition of Ricci tensor $\bar{R}_{\mu\nu}=\bar{R}^{\a}_{\mu\a\nu}=\p_{\a}\bar{\Gamma}^{\a}_{\m\n}-\p_{\n}\bar{\Gamma}^{\a}_{\m\a}+\bar{\Gamma}^{\rho}_{\m\n}\bar{\Gamma}^{\a}_{\rho\a}-\bar{\Gamma}^{\rho}_{\m\a}\bar{\Gamma}^{\a}_{\n\rho}$ and the corresponding emergent Einstein equation $\bar{E}_{\m\n}=\bar{R}_{\m\n}-\frac{1}{2}\bar{G}_{\m\n}\bar{R}=8\pi G \bar{T}_{\m\n}$ where $\bar{T}_{\m\n}=(\bar{\rho}+\bar{p})v_{\m}v_{\n}-\bar{G}_{\m\n}\bar{p}$, we can have \cite{gm7}:
\ben 
\R_{00}=-3\frac{\ddot{a}}{a}-3\frac{\dot{a}}{a}\frac{\D\,\dd}{(1-\D^2)}\equiv -3\frac{\ddot{a}}{a}+3\frac{(\ddot\phi)^{2}}{(1-\dot\phi^{2})^{2}},~
\label{23}
\een
and
\ben 
\frac{4\pi G}{3}\Big(\bar{\rho}+3\bar{p}\Big)=-\Big[\frac{\ddot{a}}{a}\frac{1}{(1-\dot\phi^{2})}+\frac{\dot{a}}{a}\frac{\dot\phi\ddot\phi}{(1-\dot\phi^{2})^{2}}\Big].
\label{24}
\een
where $\bar{\rho}$ and $\bar{p}$ are the matter density and pressure of the emergent spacetime and $G$ is the gravitational constant. 

As the {\it k}-essence model with the DBI type non-canonical Lagrangian (\ref{8}) is similar to the ideal fluid models where no vorticity occurs \cite{vikman1,vikman2}, we have utilised the energy-momentum tensor ($\bar{T}_{\m\n}$) as perfect fluid type here. The expression of 
$\bar{\rho}$ and $\bar{p}$ for flat emergent FLRW metric (\ref{21}) are \cite{gm7} $\bar{\rho}=\frac{3}{8\pi G}(\frac{\dot a}{a})^{2}\frac{1}{(1-\dot\phi^{2})}$ and $\bar{p}=\frac{-1}{8\pi G(1-\dot\phi^{2})}\Big[2\frac{\ddot a}{a}+(\frac{\dot a}{a})^{2}+2\frac{\dot a}{a}\frac{\dot\phi \ddot\phi}{(1-\dot\phi^{2})}\Big]$, where the usual matter density $\rho=\frac{3}{8\pi G}(\frac{\dot a}{a})^{2}$ and the pressure is $p=\frac{-1}{8\pi G}\Big[2\frac{\ddot a}{a}+(\frac{\dot a}{a})^{2}\Big]$.

Using Eqs. (\ref{22}) and (\ref{24}) in Eq. (\ref{23}), we have
\ben 
\bar{R}_{00}=4\pi G\Big(\bar{\rho}+3\bar{p}\Big)(1-\dot\phi^{2}).
\label{25}
\een

Therefore, Eq. \ref{20} can be written as:
\ben
\frac{d\T}{d\S}&&=\Big[-2{\sigma}^2+2{\o}^2-\frac{1}{3}\t^2-4\pi G(\bar{\r}+3\bar{p})(1-\D^2)\Big]\nonumber\\&&-\frac{7}{4}\frac{\D^2~{\dd}^2}{{(1-\D^2)}^2}+\t\frac{\D~\dd}{(1-\D^2)},
\label{26}
\een
which is {\it the required Raychaudhuri equation for timelike geodesics in the {\it{k}}-essence emergent spacetime}, where we consider the gravitational metric to be flat FLRW-type. It should be noted that if we set the {\it k}-essence scalar field $\phi$ is zero, then $\bar{\rho}$ and $\bar{p}$ reduces to the usual $\rho$ and $p$ respectively and hence Eq. (\ref{26}) reduces to the usual Raychadhuri equation as:
\ben
\frac{d\T}{d\S}=-2{\sigma}^2+2\o^2-\frac{1}{3}\t^2-4\pi G(\r+3p)\equiv\frac{d\t}{ds},
\label{27}
\een
and the corresponding Ricci tensor $\bar{R}_{\m\n}$ also reduces to the original one.

\section{Analysis of cosmological models}\label{s4}

The first thing that we are going to talk about the volume expansion $\T$ of the emergent spacetime (\ref{21}) on the basis of \cite{blau,poisson,kar,ray55,ray57,akr} and then we are going to investigate how this volume expansion $\T$ measures in a comoving Lorentz frame with $v^\a=(1,0,0,0)$. Now, based on the definition of the expansion ($\T$) cited as \cite{poisson} we obtain
\ben 
\T=D_\a v^\a&&\equiv\frac{1}{\sqrt{-\G}}~\p_\a(\sqrt{-\G}~v^\a)\nonumber\\&&
=\frac{1}{\sqrt{a^6(1-\dot\phi^2)}}\frac{d}{dt}\Big(\sqrt{a^6(1-\dot\phi^2)}\Big) \nonumber\\&&=3\frac{\dot{a}}{a}-\frac{\dot\phi\ddot\phi}{(1-\dot\phi^2)},\label{28}
\een
where $\G=\sqrt{a^6(1-\D^2)}$ is the determinant of the line element (\ref{21}). 

It's important to remember that the expanding volume of 3-space is $a^6$. According to Raychaudhuri's definition, the expansion is a measurement of the change in the transverse (cross-sectional) volume of the congruence as one travels along the geodesics, where $\delta V=\sqrt{\bar{h}}~d^{3}y$, where $\bar{h}\equiv det[\bar{h}_{ab}]$ and $\bar{h}_{ab}$ is the three-dimensional metric on the congruence's cross sections in emergent spacetime \cite{poisson}.  As the cross section grows from one location to another, $d^{3}y$ remains constant since the coordinates $y^a$ are comoving (as each geodesic travels with a constant value of its coordinates). In this case, $\T$ is proportional to the fractional rate of change in the cross-sectional volume of the congruence, denoted by $\T=\frac{1}{\delta V}\frac{d}{d\S}\delta V=\frac{1}{\sqrt{\bar{h}}}\frac{d}{d\S}\sqrt{\bar{h}}$. Because he did not expect to begin with maximally symmetric metrics on spatial slices, however, he did account for nonzero shear and rotation in his approach \cite{ray57,akr,kar,macro,Bhattacharyya1}. In this case, though, we have resorted to using $\T\equiv D_\a v^\a$ as our definition of $\T$ \cite{poisson}.

The rate of change of the scalar expansion of the {\it{k}}-essence emergent spacetime with respect to the cosmological time parameter in comoving frame ($\frac{dt}{d\S}=1$) is:
\ben
\frac{d\T}{d\S}&&=\frac{d\T}{dt}.\frac{dt}{d\S}
=3\frac{\ddot{a}}{a}-3\big(\frac{\dot{a}}{a}\big)^{2}-\Big[\frac{(\ddot\phi)^2(1+\dot\phi^{2})}{(1-\dot\phi^2)^{2}}+\frac{\dot\phi(\dddot\phi)}{(1-\dot\phi^2)}\Big].\non\\
\label{29}
\een

Now using the relation $\theta=3\Big(\frac{\dot a}{a}\Big)$ for the usual FLRW metric with the EOM relation (\ref{22}) and Eqs. (\ref{26}) and (\ref{29}), we have
\ben
\frac{\ddot{a}}{a}&&=\Big[-\frac{2}{3}\sigma^2-\frac{4\pi G}{3}(\bar{\r}+3\bar{p})(1-\dot\phi^2)+\frac{2}{3}\o^2\Big] +f(\dot\phi),~~~
\label{30}
\een
where
\ben 
f(\dot\phi)=\frac{1}{3}\frac{\dot\phi(\dddot\phi)}{(1-\dot\phi^2)}-\frac{2(\ddot\phi)^2}{{3(1-\dot\phi^2)}^2}(1+\frac{3}{8}\dot\phi^{2}).
\label{31}
\een

The given expression for $\frac{\ddot{a}}{a}$ (\ref{30}) was generated in order to investigate the cosmological singularity theorem \cite{akr} of the {\it k-}essence geometry for the FLRW background.  If we assume that all of the {\it k}-essence scalar field ($\phi$) terms in (\ref{30}) are zero, then we are able to revert to the conventional expression of the acceleration term \cite{akr} for the FLRW universe, which reads as follows: $\frac{\ddot{a}}{a}=-\frac{2}{3}\sigma^2-\frac{4\pi G}{3}(\r+3p)+\frac{2}{3}\o^2$. If we ignore the vorticity term ($\o^{2}$)  of Eq. (\ref{30}), we are able to observe that all of the terms that are contained within the third bracket are of a negative nature. This is due to the fact that $\dot\phi^{2}$ has the values $0<\dot\phi^{2}<1$. The second component contained within the third bracket is, in all actuality, the coupling term. This is the term that determines how the typical matter density and pressure are associated with the {\it k}-essence scalar field.

The question of whether or not the term $f(\dot\phi)$ is positive or negative now emerges. If $f(\dot\phi)$ is positive quantity, then the criterion of positivity of the term $f(\dot\phi)$ is:
\ben 
\dot\phi~\dddot\phi(1-\dot\phi^{2})>2(\ddot\phi)^{2}(1+\frac{3}{8}\dot\phi^{2}),
\label{32}
\een
and the condition of negativity is:
\ben 
\dot\phi~\dddot\phi(1-\dot\phi^{2})<2(\ddot\phi)^{2}(1+\frac{3}{8}\dot\phi^{2}).
\label{33}
\een

In this case, {\it we make  sure that $f(\dot\phi)\neq 0$,} since if $f(\dot\phi)= 0$, then the single contribution that the {\it k-}essence scalar fields make disappears. In order to explore the one effect that the {\it k}-essence scalar fields have in the newly emergent spacetime (\ref{21}), we need to have the necessary resources.

Now we discuss the following scenarios:

{\it Case A:} For the positivity of the term $f(\dot\phi)$, if we set 
\ben 
f(\dot\phi)>\frac{2}{3}\sigma^2+\frac{4\pi G}{3}(\bar{\r}+3\bar{p})(1-\dot\phi^2),
\label{34}
\een
then we have from (\ref{30}) $\frac{\ddot{a}}{a}>0$. That is, a {\it ``expansion"} occurs rather than collapse, indicating that the relative acceleration between two distant galaxies is growing. Therefore, we can state that we have an expanding universe if the {\it k}-essence scalar fields (\ref{31}) are the only contributors and they dominate over the shear and coupling factors.
For the condition (\ref{34}), this can be seen as a universe devoid of singularities. This goes against the big bang theory of the universe. For instance, S. G. Choudhury et al. \cite{Choudhury} illustrates that the Raychaudhuri equation may be applied to an accelerating universe in a different context.
 \\ 

{\it Case B:} On the other hand, if the term $f(\dot\phi)$ is both  negative or positive and we set 
\ben 
f(\dot\phi)<\frac{2}{3}\sigma^2+\frac{4\pi G}{3}(\bar{\r}+3\bar{p})(1-\dot\phi^2),
\label{35}
\een
then we have from (\ref{30}) $\frac{\ddot{a}}{a}<0$.

So the gravity effect causes deceleration of the expansion and this is abetted by $\frac{2}{3}\sigma^2$ and the coupled term $\frac{4\pi G}{3}(\bar{\r}+3\bar{p})(1-\dot\phi^2)$ in bringing about a {\it collapse}. So, in the absence of $\o^2$, spatial volumes inevitably collapse to singularity, which is genuinly the case about the study of standard cosmolgy in Raychaudhuri equation \cite{akr}.\\

{\it Case C:} Lastly, if we set
\ben
f(\dot\phi)=\frac{2}{3}\sigma^2+\frac{4\pi G}{3}(\bar{\r}+3\bar{p})(1-\dot\phi^2),
\label{36}
\een
then we have $\frac{\ddot{a}}{a}=0$. 

According to this stipulation (\ref{36}), a steady-state model of the cosmos is required if the relative velocity of two extremely distant galaxies is to remain unchanged which was touched on briefly in \cite{hoyle,bg,hn1,hn2,hn3,hn4}. However, one should keep in mind that the current observations  \cite{Riess,Perlmutter,Komatsu,planck1} show that the steady-state model is even now physically unacceptable.

Now we are going to determine whether the term $f(\dot\phi)$ (\ref{31}) is going to be positive or negative for some example of $\dot\phi^{2}$ as a function of time ($t>0$) while still retaining the constraint $0<\dot\phi^{2}<1$.

{\it Case-I:} Because we are interested in the negative value of the term $f(\dot\phi)$, we will use the equation 
\ben
\dot{\phi}^2(t)=\frac{1}{1+t},
\label{37}
\een 
and the plot of $f(\dot\phi)$ vs time will look like the one shown below.

\begin{figure}[H]
\centering
\includegraphics[width=6cm, height=5cm]{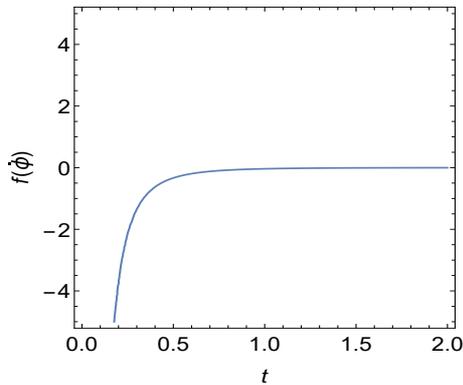}
 \caption{(Color online) Variation of $f(\dot{\phi})$ with $t$ for Case-I}\label{Fig6}
\end{figure}

When we keep the constraint $0<\dot\phi^{2}<1$ in mind, we can see from Fig. 1 that the nature of the term $f(\dot\phi)$ is such that it has a negative value. This is true for the particular choice of $\dot\phi^{2}$ (\ref{37}), which we are using.\\

{\it Case-II:} To test the positivity of $f(\dot\phi)$ while retaining the condition $0<\dot\phi^{2}<1$, we choose $\dot\phi^{2}$ as
\ben
\dot{\phi}^2(t)=\frac{1}{1+t^{0.4}},
\label{38}
\een
and the variation of $f(\dot{\phi})$ with time have shown in Fig. 2.

 \begin{figure}[H]
\centering
\includegraphics[width=6cm, height=5cm]{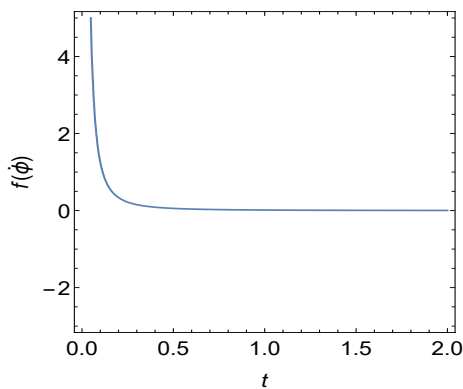}
 \caption{(Color online) Variation of $f(\dot{\phi})$ with $t$ for Case-II}\label{Fig6}
\end{figure}

This choice of $\dot{\phi}^2$ (\ref{38}) gives us positive value range of $f(\dot{\phi})$. It should be mentioned that the above two cases $f(\dot\phi)$ is going to nearly zero with respect to $t$ but not equal to zero always. 

According to the usual Raychaudhuri equation, the cosmos will collapse if the vorticity is absent, as is known by standard practise. However, despite the absence of the vorticity term, we can see that things are different here. {\it Diverse constraints (\ref{32}-\ref{36}) on the {\it k}-essence scalar field lead to different possible universe states, including collapse, expansion, and steady-state. As a result, we can infer that the {\it k}-essence scalar field imposes conditions (\ref{34},\ref{35},\ref{36}) under which all the aforementioned eventualities hold true. This spacetime is of the emergent variety (\ref{21}).} Recent observations on Supernovae type Ia and High Redshift survey, WMAP, etc. \cite{Riess,Perlmutter,Komatsu,planck1} show that the expansion of the universe has recently sped up. The {\it k}-essence theory is commonly utilised to investigate the acceleration of the present-day cosmos. However, as we can see from the literatures of \cite{picon1,picon2,picon3}, the {\it k}-essence theory is relevant both the early and the present epoch. Therefore, there is no reason to argue regarding the rationale behind why we make use of the {\it k-}essence theory to investigate the cosmic singularity by way of the Raychaudhuri equation.

\section{Focusing theorem}\label{s5}

In accordance with the Focusing theorem \cite{poisson}, an initially converging $(\t<0)$ congruence will converge more quickly in the future, and an initially diverging $(\t>0)$  congruence will diverge less quickly in the future of congruence development. So, in Einstein's GR, this theorem means that given a congruence of timelike geodesics that is hypersurface orthogonal, i.e. $\o_{\a\b}=0$, and satisfies the strong energy condition (SEC), i.e. $R_{\a\b}v^\a v^\b\geq0$, the Raychaudhuri equation predicts that $\frac{d\t}{ds}\leq0$. This implies that the growth rate should decrease as the congruence evolves and goes forward. Considering Einstein's general relativity, the strong energy condition means that gravity acts as an attractive force. In our situation, we have also complied with the SEC, as shown by the fact that $\bar{R}_{\a\b}v^\a v^\b\geq0$, which denotes that $(\bar{\rho}+3\bar{p})\geq 0$. In our particular instance, the Raychaudhuri equation (\ref{26}) tells us that $\frac{d\T}{d\bar{s}}\leq 0$ is true if we continue to hold the SEC.

However, when we consider the change that emergent scalar expansion brings about (\ref{29}), we find ourselves in a quite different scenario, i.e.,  we can no {\color{red}longer} state that $\frac{d\T}{d\bar{s}}\leq 0$. The first term ($3\frac{\ddot a}{a}$) and the last term of Eq. (\ref{29}) cannot be determined to be positive or negative. As another example, the emergent EOM (\ref{22}), reveals that we have no way of knowing whether the expression $\frac{\ddot a}{a}$ is positive or negative. 
Additionally, it is noted that the expansion $\T$ offers a naturally coordinate-independent method of determining whether the free-falling matter is converging or diverging after a certain congruence of geodesics in the gravitational field, where $\T>0$ denotes divergence and $\T<0$ denotes convergence \cite{Albareti}. In \cite{Albareti}, they have shown that in an expanding universe with accelerating expansion, the spacetime contribution to the Raychaudhuri equation is positive for the essential congruence, which is in support of a non-focusing of the congruence of geodesics.

Also, as we know from the Focusing theorem, an initially diverging congruence will diverge, whereas an initially converging congruence will converge. Therefore, depending on the {\it k}-essence scalar field terms, the rate of change of the expansion $\frac{d\T}{d\bar{s}}$ may be positive or negative. If $\frac{d\T}{d\bar{s}}>0$, then the divergence of an initially diverging congruence will be accelerated in the future, while the convergence of an initially converging congruence will be slowed down in the future \cite{Albareti,Bhattacharyya}. Now we discuss about the caustic solution based on the Eqs. (\ref{26}) and (\ref{28}).

\subsection{Caustic solutions}
Recalling the Eq.(\ref{28}) and using $\t=\frac{3\dot{a}}{a}$ as the usual scalar expansion, we get
\ben
\frac{d\T}{d\S}&&
=\frac{d\t}{dt}-\Big[\frac{(\ddot\phi)^2(1+\dot\phi^{2})}{(1-\dot\phi^2)^{2}}+\frac{\dot\phi(\dddot\phi)}{(1-\dot\phi^2)}\Big].
\label{39}
\een

Now using Eqs. (\ref{39}) and (\ref{22}) in Eq. (\ref{26}) with taking $\o^{2}=0$ we have
\ben
\frac{d\t}{dt}+\frac{1}{3}\t^2=-2\sigma^2-4\pi G(\bar{\r}+3\bar{p})(1-\dot\phi^{2})+ F(\dot\phi),\label{40}
\een
where we define
$F(\dot\phi)=\frac{\dot\phi(\dddot\phi)}{(1-\dot\phi^{2})}-\frac{2(\ddot\phi)^{2}}{{(1-\dot\phi^{2})}^2}(1+\frac{3}{8}\dot\phi^{2})$. Here, it is to be noted that we can not say whether the term $F(\dot\phi)$ is positive or negative, just like we could not say whether the term $f(\dot\phi)$ in Eq. (\ref{31}) was positive or negative. So depending on the positivity or negativity of $F(\dot\phi)$, we have discussed two different cases below.
\\

{\it Case-I:}
For positive or negative $F(\dot\phi)$ and if we consider
\ben
2\sigma^2+4\pi G(\bar{\r}+3\bar{p})(1-\D^2)\geq F(\D),
\label{41}
\een
then from Eq. (\ref{40}) we obtain
\ben
\frac{d\t}{ds}+\frac{1}{3}\t^2\leq 0,
\label{42}
\een 
where we have used $(\frac{dt}{ds}=1)$.

The above Eq. (\ref{42}) implies that
\ben
\frac{1}{\t}\geq\frac{1}{\t_0}+\frac{s}{3},
\label{43}
\een
where $\t_0$ is the value of $\t$ at $s=0$. 

The congruence of geodesics is initially convergent because the rate of change of the expansion at any given instant has a negative value. For $s\leq\frac{3}{|\t_0|}$, $\t\rightarrow -\infty$, assuming the geodesics can be stretched that far. This is a point of singular focus. As a consequence, the congruence will generate a {\it caustic} point where distinct geodesics come together and meet. This finding is important in determining the occurrence of singularities in general relativity \cite{poisson}. This is called spacetime future incompleteness or geodesic incompleteness. The scenario of a caustic situation can be depicted in Fig. 3, where $P$ is a caustic point.

\begin{figure}[H]
\centering
\includegraphics[width=4.5cm]{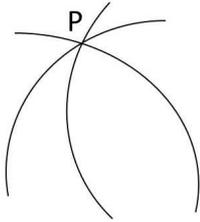}
\caption{Caustic point P}
\label{Fig :3}
\end{figure}

{\it Case II:}
For considering positive $F(\dot\phi)$ and
\ben
F(\D)>2\sigma^2+4\pi G(\bar{\r}+3\bar{p})(1-\D^2)\label{44}
\een
we get from Eq. (\ref{40}) 
\ben
\frac{d\t}{ds}+\frac{1}{3}\t^2>0.
\label{45}
\een

From this we have
\ben 
\frac{1}{\t}<\frac{1}{\t_s}+\frac{s}{3}
\label{46}
\een
where $\t_s$ is the value of $\t$ at $s=0$.

As $\frac{1}{3}\t^2$ is a positive quantity, the range of $\frac{d\t}{ds}$ (\ref{45}) is $(-\frac{1}{3}\t^2,\infty)$. So we can say that $\frac{d\t}{ds}$ could be a negative when it assigns a value $p$ for which $-\frac{1}{3}\t^2<p<0$ and it could take a positive value when $0<p<\infty$. So here arises with two conditions as:
\begin{itemize}
    \item For $-\frac{1}{3}\t^2<p<0$, $\t_s<0$, i.e., initially converging.
    \item For $0<p<\infty$, $\t_s>0$, i.e., initially diverging.
\end{itemize}

\vspace{2mm}

\begin{enumerate}
\item {\bf{Initially converging congruence $(\t_s<0)$:}}
From Eq. (\ref{46}) we get
\ben
\frac{1}{\t}<-\frac{1}{\t_s}+\frac{s}{3}.\non
\een

For $s>\frac{3}{|\t_s|}$, we have $\frac{1}{\t}<q$\, (where $q>0$). So $\frac{1}{\t}$ is less than a positive number which implies the range of $\frac{1}{\t}$ is $(q,-\infty)$. That means $\frac{1}{\t}$ passes through zero and $\t\rightarrow\infty$ but at initially $\t$ is a negative quantity, so $\t\rightarrow -\infty$. So we can say that the singularity is formed in this case.

\item {\bf{Initially diverging congruence $(\t_s>0)$:}}\\
Following Eq. (\ref{46}) and for $s>\frac{3}{|\t_s|}$, we have $\frac{1}{\t}<r$ (where $r>0$). So, again the same case is happening here as previously, but since $\t_s>0$ initially, we have always $\t\rightarrow\infty$. This means it is a {\it case of non-singularity}. So it could be a case for an {\it expanding universe}.
\end{enumerate}

The initially diverging scenario of geodesic congruences can be depicted in Fig. 4. This figure demonstrates that congruences that start out diverging invariably end up that way. 

\begin{figure}[H]
\centering
\includegraphics[width=4.5cm]{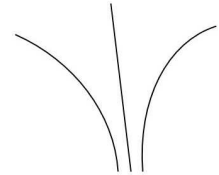}
\caption{Initially diverging congruences}
\label{Fig :4}
\end{figure}

One important thing must be noted that if we consider the equality sign of Eq. (\ref{42}), we have singularity and when $0<\frac{d\t}{ds}<\infty$,  we get a singularity-free congruence bundle of geodesics. So one can conclude that a complete divergence of geodesic congruence for initially diverging geodesics will happen if and only if there exists a finite affine parameter $s$ such that $s>\frac{3}{|\t_0|}$ and the rate of change of the expansion with respect to the finite affine parameter is always a positive quantity for which $\frac{1}{\t}<r~~(r>0)$. The above statement is only valid when Eq. (\ref{44}) holds.

Since there are conditions on whether $\frac{d\Theta}{d\bar{s}}$ is positive or not for our case, the focusing of the congruence is not obvious. If the focusing theorem does not hold true by the condition mentioned in Eqs. (\ref{34}), we can get a bouncing scenario which may be a tendency for the congruence to diverge as an accelerated expansion instead of gravitational collapse. Physically, the non-focusing exists due to huge cosmic acceleration which is large enough compared to the energy-momentum contribution of the spacetime. Due to the non-focusing theorem, there is no problem regarding future incompleteness or geodesically incomplete behaviour in the congruence of timelike geodesics. The prevention of the formation of the focusing as a semiclassical version of the singularity theorem has also been studied by Das \cite{Das} in a different context.
As a result, we may construct various types of singularity-free universe \cite{Senovilla90, Senovilla98a, Maddox}. For example in the cylindrically symmetric universe, the spacetime is completely non-singular and as a result, it is geodesically complete due to the regular behaviour of energy density and pressure everywhere \cite{Chinea92, Senovilla}. Of course, this type of classical expanding model always satisfies all energy and causality conditions, so it is a well-behaved and well-founded singularity-free universe.

Again by the condition of Eq. (\ref{44}), We have confirmed that for $\theta_s >0$ (initially diverging congruence) we have $\theta\rightarrow \infty$ within a finite affine parameter $s>\frac{3}{|\theta_s|}$. Therefore the world lines of the cosmic accelerated particles experience no singularity \cite{Bhattacharyya}. So basically divergence of the expansion scalar does imply no singularity of the congruence, but in certain cases, it leads to singularity i.e., $\theta\rightarrow -\infty$ for the initially converging congruence \cite{Mohanjan}.
Already we have mentioned in Section IV, Case-A that, a co-moving observer can experience the expansion of the congruence for the positivity of the term $f(\dot{\phi})$ (see Eq. (\ref{32}) and (\ref{34})). On the other hand, as we are dealing with cosmology, we are interested only in the evolution of scale factor $a(t)$ i.e., the acceleration or deceleration associated with the term $\frac{\ddot{a}}{a}$. If we study astrophysics, this may have some different interpretations such as there may be a supernova explosion for the case of non-focusing condition and the formation of a white dwarf, neutron star, black hole or interestingly quark star for the case of gravitational collapse. The possibility of accelerated expansion or decelerated phase underlies in Eqs. (\ref{34}) and (\ref{35}) respectively \cite{Albareti}.
	
Finally, we can state that the effect of matter on spacetime curvature causes a focusing effect (converging case) due to gravitational attraction and the defocusing (diverging case) happens due to extra interaction of k-essence scalar field $\phi$ with the usual gravity (seems to be repulsive in nature) in the congruence of timelike geodesics (see Eqs. (\ref{30}) and (\ref{31})).

The total scenario of singularity and non-singularity in one number line can be represented in the figure below (Fig. 5). In this figure, one point must be noted that singularity includes the zero point of the number line whereas non-singularity excludes the zero and tends to $+\infty$.

\begin{figure}[h]
\centering
\includegraphics[width=7.5cm]{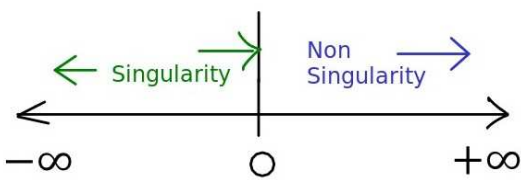}
\caption{Representation of singularity and non-singularity in one number line. }
\label{Fig:5}
\end{figure}

\section{Conclusion and Discussions}\label{s6}

As the {\it{k}}-essence scalar field model describes the entire range from the early epoch to the late time acceleration of the evolution of our universe \cite{picon1,picon2,picon3}, so in this paper, we have described the singular and non-singular nature of the universe through the Raychaudhuri equation. The non-singularity is genuinely true in the era of the late time acceleration when the universe expands. Besides this, we have dealt with collapsing universe which is a singular universe. Here, we have employed a flat FLRW-type background gravitational metric. The Focusing theorem is also explained. We identified two possible outcomes from this analysis; depending on the {\it k}-essence scalar field parameters, the expansion rate $\frac{d\T}{d\bar{s}}$ may either increase or decrease \cite{Albareti,Bhattacharyya}.

On the other hand, we have also spoken about the caustic formation approach in order to demonstrate how the {\it k-}essence Raychaudhuri equation behaves with singular and singularity-free models of the universe.  We found that all of the aforementioned possibilities are conditional, with constraints imposed by the additional interactions (forces) of {\it k}-essence scalar field coupled with the usual gravity. By the term ``conditional" we meant to indicate the condition imposed on Eq. (\ref{30}). Now if Eq. (\ref{34}) holds then from Eq. (\ref{30}) we can state that there will be a divergence of congruence (non-focusing) and if Eq.(\ref{35}) holds true, there will be a convergence of congruence (focusing) i.e., gravity effect causes deceleration of the expansion. The caustic formation is also a same kind of conditional result which has been mentioned in Eq. (\ref{41}) and (\ref{44}).

It is important to note that the {\it k-}essence model is often able to be used for the investigation of dark energy. However, the cosmic backdrop of dark energy remains relatively mysterious, and its mere existence may now be in some question \cite{Sarkar}, according to the most recent study of data from the Planck team \cite{Ade}. Here, as in Ref. \cite{gm5},  we have discussed everything from the perspective of a gravitational or geometrical theory of standpoint.

The scalar expansion or volume expansion is the rate of change of the cross-sectional area orthogonal to the bundle of geodesics. Therefore, it follows that an expansion that approaches a negative infinity suggests a convergence of the congruence, while a positive infinity implies a complete divergence. Singularity was already explained by the usual Raychaudhuri equation. A wonderful example of this kind of expanding universe, or singularity-free solution, is the {\it k}-essence scalar field model. As discussed in \cite{kar,macro}, under the assumption of homogeneity and isotropy GR predicts that the universe was in a singular state at a time less than $H^{-1}$ ago. However, the {\it k}-essence model is an exception to this rule and shows how {\it the Raychaudhuri equation may be used to get a non-singular expanding universe as well as a singular collapsing universe under certain conditions imposed by the {\it k}-essence scalar fields.}

Now we discuss the differences between the Einstein frame and the Jordan frame in our emergent geometry and its physical implications.
The main difference between the Jordan frame and the Einstein frame is, in the Jordan frame the scalar field or some function of it multiplies with the Ricci scalar whereas, in the Einstein frame in which Ricci scalar is not multiplied by the scalar field. Some of the works with non-minimal coupling of field with gravity have been done in the Jordan frame \cite{Banerjee,Nozari,Odintsov,Mukherjee}. For example, in the Jordan-Brans-Dicke theory (normally known as the Brans-Dicke theory or BD theory) the BD scalar field ($\psi$) couples non-minimally with gravity. So its corresponding action can be written in the Jordan frame \cite{Banerjee,Mukherjee}. One point must be noted that one can also write down the action of Brans-Dicke theory in the Einstein frame with the help of a conformal transformation \cite{Banerjee}, $\bar{g}_{\a\b}=\psi g_{\a\b}\nonumber$, 
while the corresponding scalar field
$\zeta=\ln \psi\nonumber$,
where $g_{\a\b}$ is the metric tensor field in the Jordan frame and the bar denotes the corresponding metric tensor in the Einstein conformal frame. Which means there exists some conformal transformation by which we can interchange between these two frames. In {\it k}-essence emergent geometry, which is a scalar field theory, we deal with the minimal coupling of {\it k}-essence scalar field ($\phi$) with gravity only. There is no term in the action of {\it k-}essence which denotes the coupling of the scalar field with Ricci scalar ($R$).

On the other hand, the Bekestein's conformally coupled scalar-tensor theory is also considered as a `non-minimally coupled scalar-tensor theory of gravity' \cite{Choudhury}. In {\it k}-essence scalar field model the Lagrangian is a function of both $X$ and $\phi$. Here $X$ denotes the coupling of the scalar field of {\it k}-essence ($\phi$) with gravity having the form \cite{vikman1},
$X=\frac{1}{2}g^{\a\b}\nabla_\a\phi\nabla_\b\phi.$
So the action of {\it k}-essence geometry is purely a scalar field model, not deals with non-minimally coupled theory or scalar-tensor theory of gravity (absence of Ricci scalar coupled with the field, unlike \cite{Nozari}). So we have not considered the action of {\it k}-essence  scalar field model in the Jordan frame, but rather in the Einstein frame. 

Consideration of the total action describing the dynamics of {\it k}-essence and general relativity in Jordan frame will be a good project in the future. It also would be a project on the study of the total action of {\it k}-essence geometry in the context of the scalar-tensor theory of gravity. The authors of the present work also would like to comment with the help of the reference \cite{Faraoni} that, the the Jordan frame formulation of the scalar-tensor theory is unstable and non-viable due to the violation of weak energy conditions as far as the classical gravity sector is concerned. If the {\it k}-essence action can be written in the Jordan frame then the result may be different from our case which we are not considering at present.\\ \\

{\bf Acknowledgement:}
SD is thankful to the International Centre for Theoretical Sciences, (ICTS-TIFR), Bengaluru, India for the hybrid program - Physics of the Early Universe (code: ICTS/peu2022/1). AP and GM acknowledge the DSTB, Government of West Bengal, India for support through the Grants No.: 322(Sanc.)/ST/P/S\&T/16G-3/2018 dated 06.03.2019. For various helpful discussions and advice, the authors additionally acknowledge Prof. Parthasarathi Majumdar, School of Physical Sciences, IACS, Kolkata and Bivash Majumder, Dept. of Mathematics, Prabhat Kumar College, Contai, W.B., India. The authors would like to thank the referees for
illuminating suggestions to improve the manuscript.\\

{\bf Keywords:}
Raychaudhuri equation, {\it k}-essence geometry, FLRW Spacetime \\ 

{\bf Conflict of Interest:}
The authors declare no conflict of interest.

\end{document}